# A first-principles study of structural, elastic, electronic, and transport properties of Cs$_2$Te

Gaoxue Wang[a], Jinlin Zhang[b], Chengkun Huang[a], Dimitre A. Dimitrov[b], Anna M. Alexander[b], and Evgenya I Simakov[b]

[a] Theoretical Division, Los Alamos National Laboratory P.O. Box 1663, Los Alamos, NM 87545, USA
[b] AOT-AE, Los Alamos National Laboratory P.O. Box 1663, Los Alamos, NM 87545, USA

**Abstract**

The pursuit to operate photocathodes at high accelerating gradients to increase brightness of electron beams is gaining interests within the accelerator community. Cesium telluride (Cs$_2$Te) is a widely used photocathode material and it is presumed to offer resilience to higher gradients because of its wider band gap compared to other semiconductors. Despite its advantages, crucial material properties of Cs$_2$Te remain largely unknown both in theory and experiments. In this study, we employ first-principles calculations to provide detailed structural, elastic, electronic and transport properties of Cs$_2$Te. It is found that Cs$_2$Te has an intrinsic mobility of 20 cm$^2$/Vs for electrons and 2.0 cm$^2$/Vs for holes at room temperature. The low mobility is primarily limited by the strong polar optical phonon scattering. Cs$_2$Te also exhibits ultralow lattice thermal conductivity of 0.2 W/(m*K) at room temperature. Based on the energy gain/loss balance under external field and electron-phonon scattering, we predict that Cs$_2$Te has a dielectric breakdown field in the range from ~60 MV/m to ~132 MV/m at room temperature dependent on the doping level of Cs$_2$Te. Our results are crucial to advance the understanding of applicability of Cs$_2$Te photocathodes for high-gradient operation.

Corresponding author:
gaoxuew@lanl.gov

# I. Introduction

In advanced photocathode community, there is a growing interest in applying high gradients to photocathodes to increase brightness of electron beams.[1, 2] One concern regarding the application of high gradients to semiconductor photocathodes is that they are susceptible to dielectric breakdown due to impact ionization.[3, 4] Impact ionization occurs when the electric field applied to a material exceeds a critical threshold, causing an abrupt increase in electrical conductivity. This phenomenon results from the acceleration of charge carriers in the material to velocities capable of creating additional free charge carriers through collisions. As a result, an avalanche process occurs, leading to a rapid increase in the number of free charge carriers and a significant decrease in the material's resistance. Such breakdown can result in failure of the device or degradation of performance.

It is reasoned that semiconducting cesium telluride ($Cs_2Te$) photocathode may be capable of withstanding higher electric fields due to the larger band gap of 3.3 eV in $Cs_2Te$ compared to other semiconducting materials, such as alkali antimonide. $Cs_2Te$ is a popular photocathode material because of its high quantum efficiency (QE) exceeding 10% at its working wavelength, long lifetimes over a span of several months, stability under various operating conditions, and relatively fast response times.[5] $Cs_2Te$ photocathodes are in operational photoinjectors[6] for several major accelerator facilities such as LCLS-II (SLAC, USA), FLASH (DESY), and European XFEL (Germany), as well as in many smaller accelerators and test facilities worldwide.[7] At Los Alamos National Laboratory (LANL), we selected to use $Cs_2Te$ to generate ultra-bright electron beam at the Cathodes and Radiofrequency Interactions in Extremes (CARIE) photocathode test stand.[8]

Despite its widespread use, many material properties of $Cs_2Te$ relevant to its interactions with strong electromagnetic fields remain unknown. For instance, to the best of our knowledge, the electrical and thermal conductivities of $Cs_2Te$ have never been reported in the literature. The electrical conductivity is crucial for understanding photocathode behavior, as poor electrical conductivity may result in ionized surfaces, impacting photoemission or even integrity of the photocathode.[9] Thermal conductivity is essential for heat dissipation during operation of photocathodes. Furthermore, the performance of $Cs_2Te$ photocathodes under high gradients remains unclear and requires detailed understanding of the transport properties. Additionally, one quantity that is absolutely necessary to study is the dielectric breakdown field that $Cs_2Te$ photocathodes can withstand and its dependence on RF frequency of the electromagnetic field.

In this study, we employ first-principles calculations based on density functional theory (DFT) and density functional perturbation theory (DFPT) to predict fundamental properties of $Cs_2Te$, including structural, elastic, vibrational, electronic and transport properties. We compute the electron-phonon and phonon-phonon scattering rates of $Cs_2Te$ for the first time using first-principles approach, enabling the calculations of electron and phonon transport via the Boltzmann transport equation (BTE). Our calculations reveal that $Cs_2Te$ has an ultralow thermal conductivity of 0.2 W/(m*K) due to its low group velocity, indicating that heat dissipation in $Cs_2Te$ photocathodes will be inefficient, potentially impacting high power applications. We predict room-temperature intrinsic mobilities of 20 cm$^2$/Vs for electrons and 1.2 cm$^2$/Vs for holes, primarily limited by the strong polar optical phonon scattering.

Additionally, using the von Hippel low-energy criterion[10] the electron-phonon scattering rates allow us to estimate that the breakdown field varies in between 60 MV/m and 132 MV/m at room temperature depending on the doping level of $Cs_2Te$. These calculated results provide critical inputs for modeling $Cs_2Te$ photocathodes, particularly regarding detailed photoemissive properties based on Monte Carlo simulations,[11-13], which are key to advance understanding of applicability of $Cs_2Te$ photocathodes in high-gradient accelerators.

## II. Crystal structure of $Cs_2Te$

To determine the crystal structure and atomic positions in $Cs_2Te$, we fully relaxed the lattice parameters and atomic coordinates using the Vienna ab-initio simulation package (VASP) [1, 2] based on DFT. The Perdew-Burke-Ernzerhof (PBE) [14] functional of the generalized gradient approximation (GGA) was used for the electron exchange-correlation. Standard PAW-PBE pseudopotentials [15] were used with a cutoff energy of 500 eV for the plane waves. The structure was converged with the maximum force on each atom smaller than 0.01 eV/Å. $Cs_2Te$ belongs to the Pnma space group (no. 62) with an orthorhombic unit cell characterized by the three lattice vectors a, b, and c. The relaxation yielded lattice constants of a=9.62, b=5.89, c=11.63 Å. The comparison between the computed and measured lattice constants is shown in Table 1 [16, 17]. In $Cs_2Te$, Cs atoms occupy two inequivalent sites with different coordination and bond character, denoted as $Cs_I$ and $Cs_{II}$ in Figure 1(a). $Cs_I$ is bonded with 5 neighboring Te atoms with bond length smaller than 4.5 Å. $Cs_{II}$ is bonded with 4 Te atoms with a bond length smaller than 3.9 Å. The average Cs-Te bond length in $Cs_2Te$ is approximately 3.95 Å.

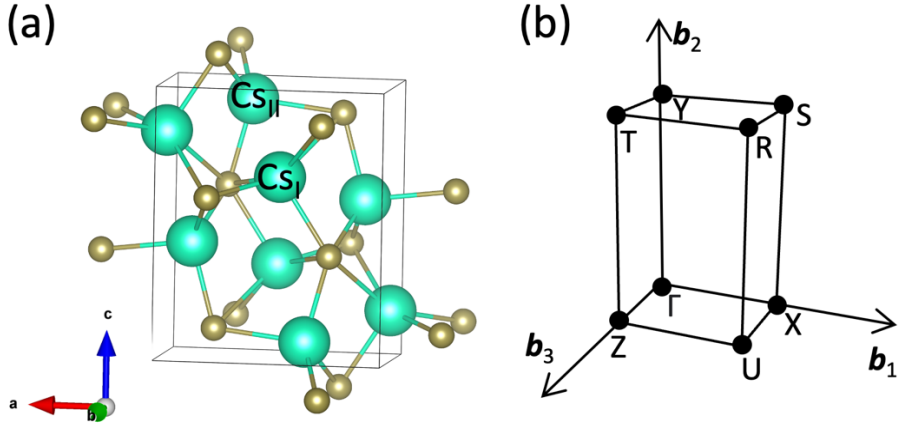

Figure 1. (a) Crystal structure of $Cs_2Te$ where the larger atoms are Cs and the smaller atoms are Te. $Cs_I$ and $Cs_{II}$ denote Cs atoms that occupy two inequivalent sites. (b) the schematic of the Brillouin zone used in calculating the electronic band structure and the phonon dispersion. Rendered using VESTA.[18]

Table 1. Calculated and experimental lattice constants of $Cs_2Te$.

|   | This work | Experiment [16, 17] |
|---|---|---|
| a | 9.62 | 9.512 |
| b | 5.89 | 5.838 |
| c | 11.63 | 11.748 |

## III. Elastic properties of Cs$_2$Te

Elastic properties are crucial to know in numerous technological applications. They influence the fracture, toughness, and sound velocities in the material. While a crystal theoretically has 36 independent elastic constants, the inherent symmetry of the crystal significantly reduces the number of independent elastic constants. Cs$_2$Te has only eight independent elastic constants: $C_{11}$, $C_{12}$, $C_{13}$, $C_{22}$, $C_{33}$, $C_{44}$, $C_{55}$ and $C_{66}$. These elastic constants can be averaged by integration on the unit sphere to calculate the standard Young modulus, bulk modulus, shear modulus, and Poisson's ratio. Utilizing the ELATE code,[19] we computed the elastic properties of Cs$_2$Te, which are presented in Table 2. The elastic constants clearly satisfy the Born stability criteria [20] for materials with orthorhombic structures: (i) $C_{11} > 0$; (ii) $C_{11}*C_{22} > C_{12}^2$; iii) $C_{11}*C_{22}*C_{33} + 2C_{12}*C_{13}*C_{23} - C_{11}*C_{23}^2 - C_{22}*C_{13}^2 - C_{33}*C_{12}^2 > 0$; (iv) $C_{44} > 0$; (v) $C_{55} > 0$; (vi) $C_{66} > 0$. The calculated bulk modulus and shear modulus are 9.52 GPa and 3.72 GPa, respectively. The small bulk and shear moduli indicate Cs$_2$Te is a relatively soft material with weak chemical bonds, likely due to the large Cs-Te bond length as described in the previous section.

Table 2. Calculated elastic constants, bulk modulus $B$, shear modulus G, Young's modulus E, Poisson's ratio ν of Cs$_2$Te. The values reported in this table are from the Void-Reuss-Hill averaging approach.[21-23]

| $C_{11}$ (GPa) | $C_{12}$ (GPa) | $C_{13}$ (GPa) | $C_{22}$ (GPa) | $C_{33}$ (GPa) | $C_{44}$ (GPa) | $C_{55}$ (GPa) | $C_{66}$ (GPa) | B (GPa) | G (GPa) | E (GPa) | ν |
|---|---|---|---|---|---|---|---|---|---|---|---|
| 12.83 | 6.65 | 6.64 | 14.17 | 18.21 | 2.20 | 4.16 | 4.95 | 9.52 | 3.72 | 9.88 | 0.33 |

## IV. Vibrational properties of Cs$_2$Te

We investigated the vibrational properties of Cs$_2$Te using density functional perturbation theory (DFPT). To ensure the convergence of the vibrational frequencies, we constructed a 1×3×1 supercell for phonon dispersion calculations. A Monkhorst–Pack k-mesh of 2×1×2 was used to sample the Brillouin zone (BZ). Harmonic second-order interatomic force constants (IFCs) were derived using the finite displacement method, utilizing force constants obtained from VASP calculations. Subsequently, we calculated the phonon dispersion of Cs$_2$Te using the PHONOPY package based on these IFCs.[24] Our calculated phonon dispersion, shown in Figure 2, reveals absence of imaginary frequencies, indicating the dynamical stability of the structure. The highest vibrational frequency is observed to be approximately 100 cm$^{-1}$ at the Γ-point. Moreover, a discontinuity in the phonon dispersion near the Γ-point is evident, which is caused by the anisotropy of crystal structure. In our calculations, we considered the non-analytic term corrections, addressing the influence of the polarization field on the optical phonon behavior around the Γ-point. The strong anisotropy of the polarization along different directions leads to direction-dependent corrections, resulting in the discontinuity of the phonon dispersion at the Γ-point when the phonon wave vector changes direction.

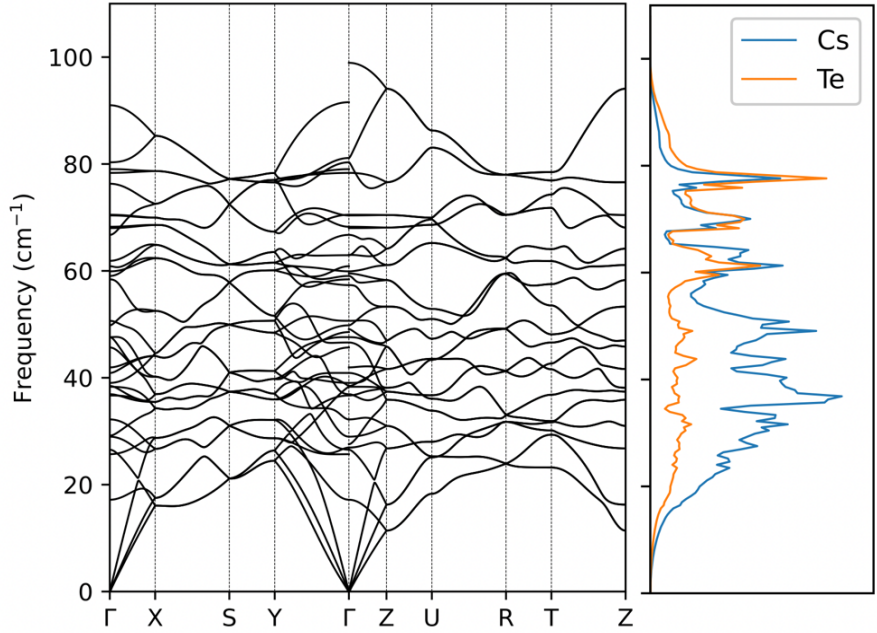

Figure 2. Phonon dispersion and phonon density of states of Cs$_2$Te.

The vibrational frequencies of each phonon mode at the Γ-point are listed in Table 3. The Raman and Infrared activity of each mode are also shown in the Table. There are 18 Raman active mode (6Ag+3B1g+6B2g+3B3g) and 12 Infrared active modes (5B1u+2B2u+5B3u). These computed vibrational frequencies can be compared with experimental Raman/Infrared spectra to validate the quality of the grown Cs$_2$Te films.

Table 3. Phonon frequencies of Cs$_2$Te at the center of Brillouin zone (Γ-point), and their Raman/Infrared activity. The three zero-frequency acoustic modes are not included in the table.

| Mode | Frequency (cm$^{-1}$) | Activity |
| --- | --- | --- |
| Au  | 17.18 | - |
| B1u | 25.70 | Infrared |
| Ag  | 26.54 | Raman |
| B3g | 28.98 | Raman |
| B3u | 29.18 | Infrared |
| B1g | 32.21 | Raman |
| Ag  | 36.69 | Raman |
| B1g | 36.92 | Raman |
| B2g | 38.31 | Raman |
| Au  | 38.47 | - |
| Ag  | 39.04 | Raman |
| B3g | 40.90 | Raman |
| B2u | 41.94 | Infrared |
| B1u | 45.67 | Infrared |
| B3u | 46.60 | Infrared |
| B2g | 47.60 | Raman |
| Ag | 49.82 | Raman |
| B2g | 50.66 | Raman |
| B1u | 58.42 | Infrared |
| B2g | 59.06 | Raman |
| Ag | 59.92 | Raman |
| B1u | 60.87 | Infrared |
| B3u | 61.88 | Infrared |
| B2g | 66.74 | Raman |
| B2u | 68.01 | Infrared |
| Au | 68.26 | - |
| B3g | 70.37 | Raman |
| B1g | 70.51 | Raman |
| B3u | 76.31 | Infrared |
| Ag | 78.34 | Raman |
| B1u | 79.03 | Infrared |
| B2g | 80.32 | Raman |
| B3u | 90.99 | Infrared |

## V. Lattice thermal conductivity of Cs$_2$Te

Thermal conductivity plays a pivotal role in managing heat within high-power devices and is of course important to know for managing heat in photocathodes. To compute lattice thermal conductivity, we used the anharmonic third-order IFCs. Utilizing the same 1×3×1 supercell and 2×1×2 Monkhorst–Pack k-mesh as in Section IV, we got these anharmonic third-order IFCs, considering interactions up to fourth nearest neighbors. Additionally, we obtained the dielectric tensor and Born effective charges based on DFPT to account for long-range electrostatic interactions. The high-frequency dielectric tensor exhibits slight anisotropy with $\varepsilon_{xx}$ = 4.09, $\varepsilon_{yy}$ = 4.06, and $\varepsilon_{zz}$ = 4.13. The computed diagonal Born effective charges are (1.21, 1.30, 1.15) for Cs$_I$, (1.04, 0.96, 1.12) for Cs$_{II}$, and (-2.26, -2.25, -2.28) for Te in units of electron charge. The off-diagonal components are lower than 0.1 and not reported here. The positive Born effective charge for Cs atoms and the negative Born effective charge clearly indicate the ionic nature of the bonds in Cs$_2$Te. The intrinsic lattice thermal conductivity can be calculated by solving the phonon BTE as implemented in ShengBTE code[25] that considers the phonon-phonon scattering processes.

The calculated lattice thermal conductivity of Cs$_2$Te is shown in Figure 3(a), revealing slight anisotropy along different directions. As temperature increases, phonon scattering intensifies, leading to a decrease in lattice thermal conductivity. At 300 K, the average thermal conductivity of Cs$_2$Te is predicted to be ~0.2 W/(m*K), calculated as the average of thermal conductivities along the three crystallographic axes. This low lattice conductivity places Cs$_2$Te among the materials with ultralow lattice thermal conductivity.[26] In comparison, Cu photocathodes have a thermal conductivity of ~401 W/(m*K) at room temperature.[27] The low thermal conductivity of Cs$_2$Te photocathodes implies that heat generated in photocathodes by either illuminating lasers or joule heating, cannot be efficiently dissipated. This deficiency can lead to localized heating and vibrations that may affect quality of generated electron beams. Thus, active cooling methods may be required when operating Cs$_2$Te photocathodes under high gradients.

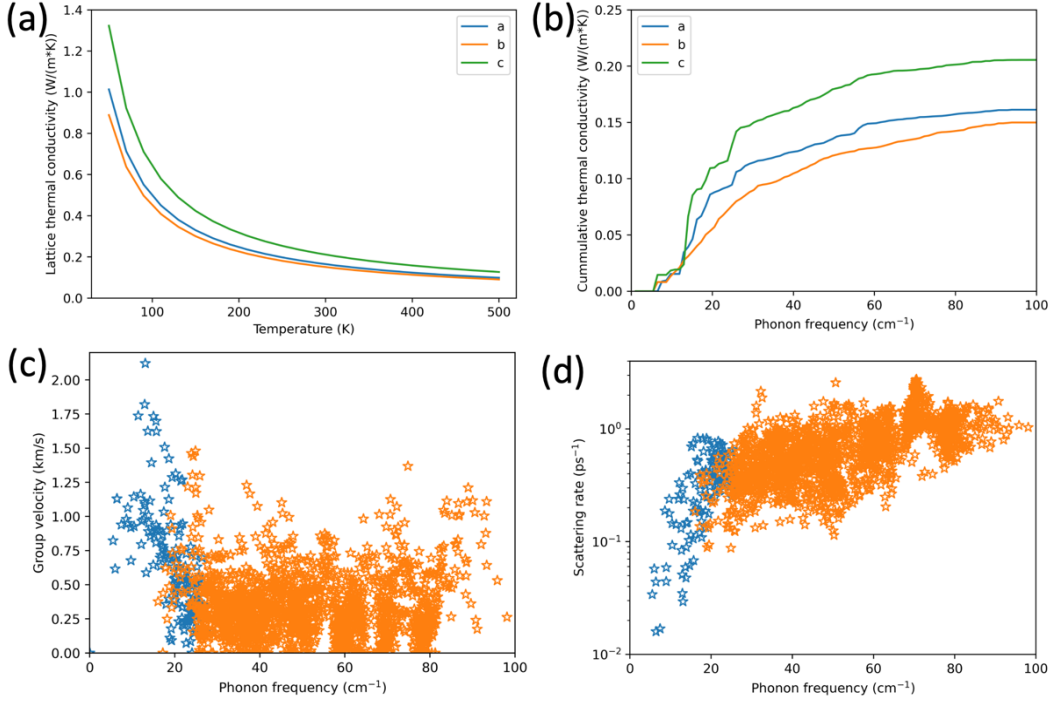

Figure 3. (a) Lattice thermal conductivity, (b) cumulative lattice thermal conductivity, (c) group velocity, and (d) phonon scattering rate of Cs$_2$Te at 300 K. The acoustic and optical modes are illustrated with blue and orange stars, respectively.

The cumulative thermal conductivity as a function of phonon vibrational frequency of Cs$_2$Te at 300 K is shown in Figure 3(b). It is seen that phonons with frequencies below 20 cm$^{-1}$, namely the acoustic phonons, contribute to nearly 50% of the lattice thermal conductivity. In the frequency range of 20-40 cm$^{-1}$, where acoustic and optical phonons interact, more than 20% of thermal conductivity is contributed. Therefore, the thermal conductivity of Cs$_2$Te is mainly determined by acoustic phonons and low frequency optical phonons. The maximum mean free path (MFP) of phonons is close to 100 nm and phonons with MFP less than 30 nm contribute 70% to the thermal conductivity. To further clarify the microscopic mechanism behind the ultralow lattice thermal conductivity for Cs$_2$Te, we plotted the group velocity and phonon-phonon scattering rates (Figures 3(c, d)). The plots show that low-frequency acoustic phonons exhibit higher group velocity and lower scattering rates, while optical phonons have high scattering rates and low group velocities, thus contributing less to thermal conductivity. The overall low lattice thermal conductivity results from the heavy atomic mass of Cs, and the relatively weak chemical bonds between Cs and Te as seen from the small bulk and shear moduli of Cs$_2$Te in Section III.

# VI. Electronic structure of Cs$_2$Te

The calculated band structure of Cs$_2$Te with GGA-PBE functional is shown in Figure 4(a). The calculated band gap is ~1.7 eV, notably lower than the experimental value of ~3.3 eV.[28] Our results align with previous DFT calculations.[29] It is acknowledged that the GGA-PBE functional tends to underestimate band gaps of semiconducting materials.[30] However, despite this limitation, the band dispersion obtained with GGA-PBE functional is similar to that calculated from more advanced functionals, such the HSE06 hybrid functional.[29] Therefore, we anticipate that GGA-PBE functional accurately captures the effective masses and group velocities of electrons and holes, which determine their transport properties. From the band dispersion, the calculated effective masses for electrons are $0.267m_0$, $0.270m_0$, and $0.267m_0$ along the Γ-X, Γ-Y, and Γ-Z directions, respectively, where $m_0$ is the rest mass of an electron. Much larger effective masses are observed for holes, which are $-2.524m_0$, $-0.732m_0$, $-1.900m_0$, along the Γ-X, Γ-Y, and Γ-Z directions, respectively.

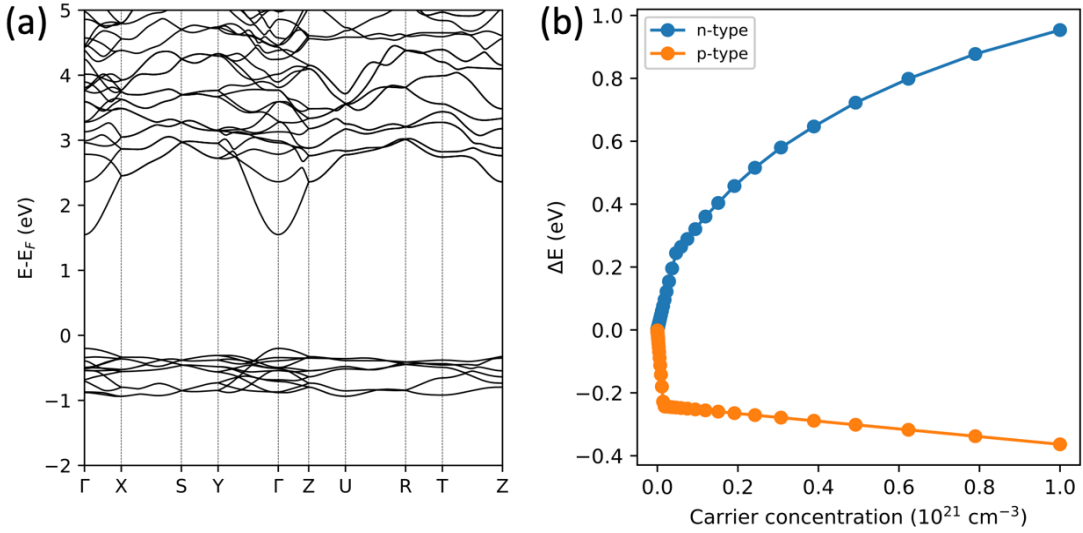

Figure 4. (a) Electronic band structure, and (b) Burstein-Moss energy shift as a function of carrier concentration of Cs$_2$Te. Positive shifts for electrons, and negative shifts for holes.

The change of the Fermi level as a function of carrier concentration due to Burstein-Moss effect can be calculated using[31, 32]

$$\Delta E = \frac{h^2}{8\pi^2 m^*}(3\pi^2 n)^{2/3}, \quad (1)$$

where $h$ is Planck's constant, and $n$ is the concentration of charge carriers (electrons or holes) associated with the doping level of semiconductors. The energy shifts caused by excess electrons are calculated relative to the conduction band minimum (CBM), and those induced by excess holes are referenced to the valance band maximum (VBM). Figure 4(b) depicts the shifts of the Fermi level as a function of carrier concentration of electrons (n-type) and holes (p-type) in Cs$_2$Te. With an electron concentration of $10^{21}$ cm$^{-3}$, the Fermi level is observed to shift 0.8 eV above CBM, whereas with a same hole concentration, it shifts only 0.4 eV below VBM. This difference arises from the narrower dispersion of the valance bands compared to the conduction bands, as evidenced by the band structure. The transport properties of electrons and holes are evaluated within the carrier concentration range of $10^{18}$ cm$^{-3}$ (slightly doped Cs$_2$Te) to $10^{21}$ cm$^{-3}$ (heavily doped Cs$_2$Te).

## VII. Electron-phonon scattering and carrier mobility in Cs$_2$Te

In order to gain fundamental understanding of the electron transport properties in Cs$_2$Te at high electromagnetic fields, we investigated the electron-phonon scattering. We calculated the electron transport properties with the ab-initio scattering and the transport method implemented in the AMSET code.[33, 34] The package solves the electronic BTE in the momentum relaxation-time approximation (MRTA) to calculate scattering rates and mobilities within the Born approximation. In AMSET code, the relaxation time is estimated by the number of different band and *k*-point dependent scattering rates instead of the constant relaxation-time approach. In this work, we calculated scattering rates by considering the following scattering processes: (a) acoustic deformation potential (ADP), which is responsible for the phonon-electron interactions; (b) ionized impurity (IMP) scattering, which represents scattering of charge carriers by ionization of the lattice; and (c) polar optical phonon (POP) scattering, which considers the interaction between polar optical phonons and electrons. The resulting carrier relaxation time, namely the inverse of the scattering rate $\tau$, was calculated by Matthiessen's rule[33]

$$\frac{1}{\tau} = \frac{1}{\tau^{ADP}} + \frac{1}{\tau^{IMP}} + \frac{1}{\tau^{POP}}, \qquad (2)$$

where $\tau^{ADP}$, $\tau^{IMP}$, and $\tau^{POP}$ are the relaxation times from the ADP, IMP, and POP scattering, respectively. In ADP and IMP scatterings, electrons do not acquire or lose energy while undergoing elastic scattering, whereas POP scatterings exhibit inelastic scattering that occurs due to phonon emission or absorption. The Fermi's golden rule is used to calculate the scattering rates for elastic and inelastic scattering from an initial |*nk*> state to the final |*mk+q*> state. The scattering rate can be written as

$$\tau^{-1}_{n\mathbf{k} \to m\mathbf{k}+\mathbf{q}} = \frac{2\pi}{\hbar} |g_{nm}|^2 \delta(\epsilon_{n\mathbf{k}} - \epsilon_{m\mathbf{k}+\mathbf{q}}), \qquad (3)$$

where $\epsilon_{n\mathbf{k}}$ and $\epsilon_{m\mathbf{k}+\mathbf{q}}$ are the energy of the initial state and the final state after scattering, $g_{nm}$ is the electron-phonon coupling matrix element that is defined as

$$g_{nm}(\mathbf{k}, \mathbf{q}) = \langle m\mathbf{k} + \mathbf{q} | \Delta_q V | n\mathbf{k} \rangle, \qquad (4)$$

where $\Delta_q V$ is the electronic perturbation from different scattering mechanisms. The required parameters to obtain the scattering rates, such as dense and uniform band structures, elastic coefficients, wave-function coefficients, deformation potentials, static and high-frequency dielectric constants, and polar-phonon frequency, were calculated from DFT and DFPT.

The calculated electron and hole scattering rates at 300 K are shown in Figures 5(a, b). The figures show that the POP scattering clearly dominates due to the pronounced polarity of the chemical bonds in Cs$_2$Te. The ADP scattering and IMP scattering rates have much smaller contributions to the total scattering rates. The electron scattering rate is around 450 ps$^{-1}$ near the CBM (Figure 5(b)). The hole scattering rate at VBM is around 850 ps$^{-1}$. These calculated scattering rates were used to calculate the carrier transport and to predict the breakdown field of Cs$_2$Te.

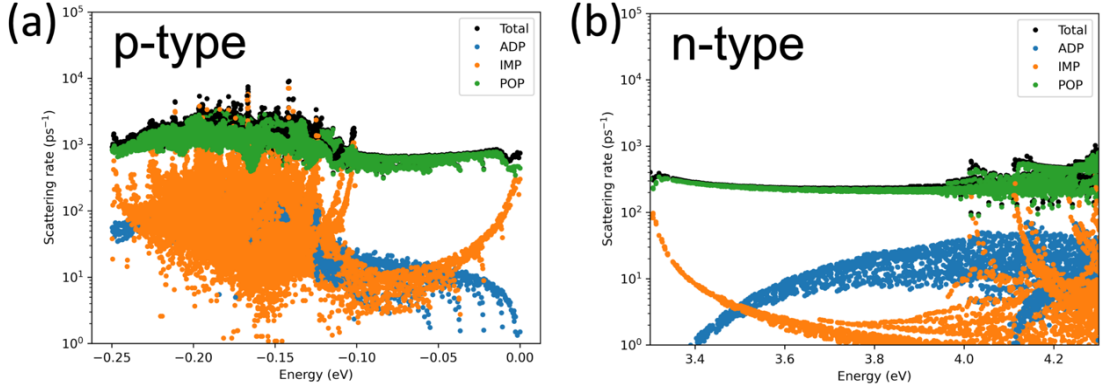

Figure 5. Electron-phonon scattering rates for (a) p-type carrier (holes), and (b) n-type carrier (electrons) of Cs$_2$Te at 300 K. The carrier concentration is $10^{18}$ cm$^{-3}$. Black dots are the total scattering rates. Green, orange, and blue dots represent the contribution from POP, ADP, and IMP scattering, respectively. The Fermi level is set to 0 eV. Negative energy in (a) is related to holes, and positive energy in (b) is related to electrons. A scissor correction is added to shift the CBM to experimental value of 3.3 eV.

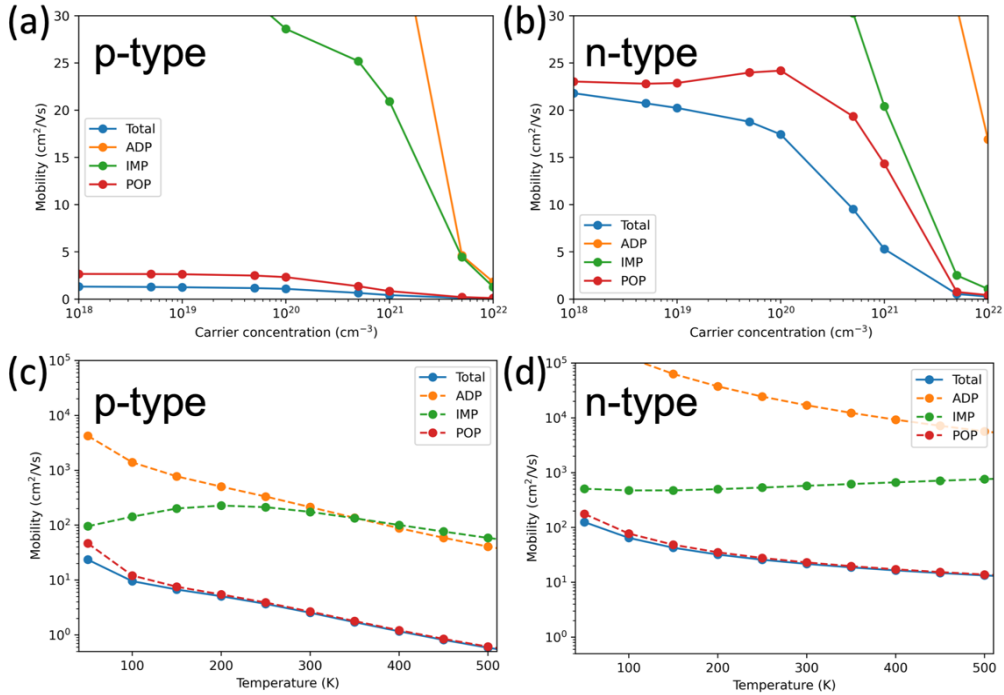

Figure 6. (a, b) Carrier mobility as a function of carrier concentration for p-type (holes) and n-type (electrons) carriers in Cs$_2$Te at 300 K. (c, d) Carrier mobility as a function of temperature with a carrier concentration of $10^{18}$ cm$^{-3}$.

The calculated hole and electron mobilities as functions of temperature and carrier concentrations are shown in Figure 6. In the slightly doped Cs$_2$Te with an electron concentration of $10^{18}$ cm$^{-3}$, the hole mobility is ~2.0 cm$^2$/(V*s) at 300 K (Figure 6(a)), and the electron mobility is around 20 cm$^2$/(V*s) (Figure 6(b)). Both electron and hole mobilities slightly decrease with carrier concentrations up to $10^{20}$ cm$^{-3}$, and rapidly decrease when the carrier concentration is larger than $10^{20}$ cm$^{-3}$. Hole and electron mobilities also decrease with temperature as seen from Figures 6(c, d), due to the increased scattering rates at higher temperatures. The much lower mobility of holes than that of electrons is related to the larger scattering rates and the large effective masses of holes.

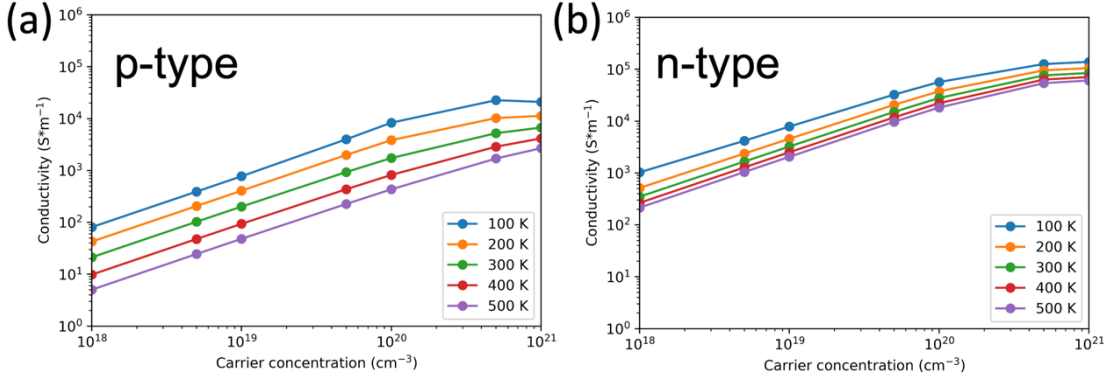

Figure 7. Electrical conductivity of (a) p-type and (b) n-type Cs$_2$Te as a function of carrier concentrations at different temperatures.

Figures 7(a, b) show calculated electrical conductivities of p-type and n-type doped Cs$_2$Te. Both p-type and n-type conductivities increase monotonously with carrier concentration from $10^{18}$ cm$^{-3}$ to $10^{21}$ cm$^{-3}$. The increase of conductivities with carrier concentrations is due to the increase in the number of carriers participating in the conduction process. The electrical conductivities decrease with temperature, which can be attributed to the increased electron-phonon scattering rates at higher temperatures. Moreover, it is seen that the n-type conductivity is around one order of magnitude larger than the p-type conductivity at the same carrier concentration and temperature.

## VIII. Dielectric breakdown in Cs$_2$Te

The dielectric breakdown field of Cs$_2$Te is estimated via a fully first-principles approach based on the Frohlich-von Hippel criterion.[10, 35, 36] This method contemplates that an electron avalanche occurs when the electron energy reaches the threshold to excite a valence electron across the electronic energy gap, causing electron multiplication and leading to impact ionization. The dielectric breakdown occurs when the rate of energy gain $A(E,\epsilon,T)$ of an electron with energy $\epsilon$ due to the external field $E$ at temperature $T$ is larger than the energy-loss rate $B(\epsilon,T)$ due to electron-phonon scattering

$$A(E,\epsilon,T) > B(\epsilon,T), \tag{5}$$

for energies $\epsilon$ ranging from CBM to the impact ionization threshold of CBM+E$_g$, where E$_g$ is the band gap of the material.[10, 36-38] The energy-gain rate of the electron can be evaluated as

$$A(E,\epsilon,T) = \frac{1}{3}\frac{e^2\tau(\epsilon,T)}{m^*}E^2, \tag{6}$$

where $e$ and $m^*$ are the electronic charge and effective mass, respectively. $\tau(\epsilon, T)$ is the energy- and temperature-dependent electron-phonon scattering rate, as shown in Figure 5(b). The field-independent rate of energy loss $B(\epsilon, T)$ to the lattice is obtained by subtracting the rate of phonon absorption from phonon emission[36]

$$B(\epsilon, T) = \frac{2\pi}{\hbar D(\epsilon)} \sum_{\pm} \sum_{n\mathbf{k}} \sum_{m\mathbf{k}+\mathbf{q}} \left\{ \begin{array}{l} \pm \hbar \omega_q |g_{nm}(\mathbf{k}, \mathbf{q})|^2 \left(n_q + \frac{1}{2} \mp \frac{1}{2}\right) \\ \times \delta(\epsilon_{n\mathbf{k}} - \epsilon_{m\mathbf{k}+\mathbf{q}} \pm \hbar \omega_q) \delta(\epsilon_{n\mathbf{k}} - \epsilon) \end{array} \right\}, \qquad (7)$$

where $n_q$ is the phonon occupation number that is given by Bose-Einstein distribution, $\pm$ sign represents the adsorption and emission of phonon, respectively, and $\omega_q$ is the phonon frequency. We used an effective phonon frequency that is defined as the weighted sum over all phonon modes at Γ-point.[33] In Figure 8(a) we present the energy loss rate as a function of energy (black line) and energy gain rate under different external fields (blue lines), starting from the CBM. The intrinsic breakdown field is defined as the smallest external electric field such that the energy gain rate is larger than the energy loss rate for all energies between the CBM and the CBM plus the band gap (3.3 eV). We found that that the breakdown field is ~60 MV/m at room temperature for $Cs_2Te$ with an electron concentration of $10^{18}$ cm$^{-3}$. Using the same approach, we obtained the breakdown fields for different carrier concentrations and temperatures as shown in Figure 8(b). The breakdown field clearly increases with temperature and carrier concentrations due to the increased energy loss rate at higher temperatures and higher carrier concentrations. The dielectric breakdown field of $Cs_2Te$ at 300 K is predicted to vary from 60 MV/m to 132 MV/m depending on the doping level of samples. It is to be noted that the von Hippel low-energy criterion should be seen as a lower bound for the breakdown field. This is because $Cs_2Te$ is not available in single-crystalline form since it is grown via deposition methods that give rise to polycrystalline samples with varying grain sizes. The scattering introduced by the grain boundaries may further contribute to energy loss rate and increase the breakdown field. Additionally, the method considers the breakdown due to an applied DC field, which may not be suitable in conditions in RF photoinjectors where photocathodes are exposed to RF fields.

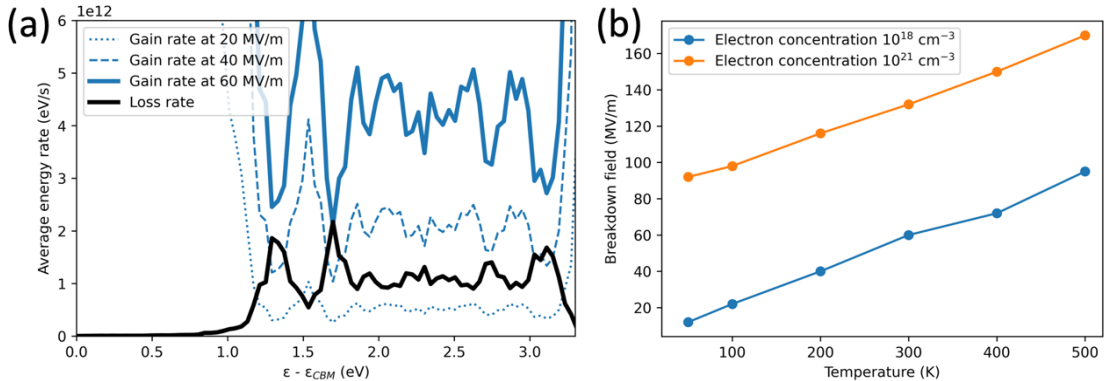

Figure 8. Dielectric breakdown of $Cs_2Te$. (a) Energy-gain rate $A(E, \epsilon, T)$ from an applied external field and energy loss rate $B(\epsilon, T)$ due to electron-phonon scattering. Both quantities are calculated at 300 K for an electron concentration of $10^{18}$ cm$^{-3}$. The intrinsic breakdown occurs when the applied electric field is such that the energy gain rate is larger than the loss rate for all energies between the conduction-band minimum (CBM) and the CBM plus the energy of the band gap (3.3 eV). (b) The dielectric breakdown field of $Cs_2Te$ as a function of temperature for slightly ($10^{18}$ cm$^{-3}$) and heavily ($10^{21}$ cm$^{-3}$) doped $Cs_2Te$.

## IX. Conclusion

This paper reports results of our investigation of the structural, elastic, electronic, vibrational, and transport properties of $Cs_2Te$ using first-principles calculations. We computed the electron-phonon and phonon-phonon scattering rates of $Cs_2Te$ using first-principles approach for the first time, enabling calculations of electron and phonon transport via the Boltzmann transport equation (BTE). We computed material properties of $Cs_2Te$, such as lattice thermal conductivity, carrier mobilities and conductivities, based on the electron-phonon and phonon-phonon scattering. We predicted the intrinsic electron and hole mobility of $Cs_2Te$ to be 20 and 1.2 cm$^2$/Vs, respectively, at room temperature. The low mobility is explained by the strong polar optical phonon scattering. We predicted that $Cs_2Te$ has ultralow lattice thermal conductivity of 0.2 W/(m*K) at room temperature due to low group velocities and relatively weak bonds in $Cs_2Te$. Based on the electron-phonon scattering rates and the von Hippel low-energy criterion, the dielectric breakdown field of $Cs_2Te$ is predicted to vary between 60 MV/m 132 MV/m at room temperature, depending on the doping of the material. Our studies serve as a basis for understanding the high-gradient performance of $Cs_2Te$ photocathodes.

## Acknowledgements

The authors gratefully acknowledge the funding for this project from the Laboratory Directed Research and Development program of Los Alamos National Laboratory (LANL) under projects 20230011DR and 20230808ER. LANL is operated by Triad National Security, LLC, for the National Nuclear Security Administration of U.S. Department of Energy (Contract No. 89233218CNA000001). This research used resources provided by the Los Alamos National Laboratory Institutional Computing Program.